\documentstyle[11pt]{article}
\newcommand{\beq}{\begin{equation}}
\newcommand{\eeq}{\end{equation}}
\newcommand{\beqr}{\begin{eqnarray}}
\newcommand{\eeqr}{\end{eqnarray}}

\newcommand{\sss}{\vspace{.2in}}
\newcommand{\sms}{\vspace{.1in}}
\def\rf#1{(\ref{eq:#1})}
\def\lab#1{\label{eq:#1}}
\def\nonu{\nonumber}
\def\br{\begin{eqnarray}}
\def\er{\end{eqnarray}}
\def\be{\begin{equation}}
\def\ee{\end{equation}}

\def\Blb{\Bigl\lbrack}
\def\Brb{\Bigr\rbrack}

\def\({\left(}
\def\){\right)}

\def\l{\lambda}

\def\om{\omega}

\def\cM{{\cal M}}

\newcommand\fourmat[4]{\left(\begin{array}{cc}  
{#1} & {#2} \\ {#3} & {#4} \end{array} \right)}

\begin{document}
~\hfill{\footnotesize UICHEP-TH/97-6; IP/BBSR-97-23}\\
\sss
\begin{center}
{\Large {\Large \bf Cyclic Shape Invariant Potentials }}
\end{center}
\vspace{.5in}
\begin{center}
{\large{\bf Uday P. Sukhatme$^{a,}$\footnote{sukhatme@uic.edu},
Constantin Rasinariu$^{a,}$\footnote{costel@uic.edu} and 
Avinash Khare$^{b,}$\footnote{khare@iop.ren.nic.in} }}
\end{center}
\vspace{.6in}
\noindent
a) \hspace*{.2in}
Department of Physics (m/c 273), University of Illinois at Chicago,\\
\hspace*{.4in}845 W. Taylor Street, Chicago, Illinois 60607-7059,USA \\
b) \hspace*{.2in}
Institute of Physics, Sachivalaya Marg, Bhubaneswar 751005, India \\
\vspace*{0.8in}
\begin{abstract}
  We formulate and study the set of coupled nonlinear differential 
equations which define a series of shape invariant potentials which repeats 
after a cycle of $p$ iterations. These cyclic shape invariant 
potentials enlarge the limited reservoir of known analytically 
solvable quantum mechanical eigenvalue problems. At large values of $x$, 
cyclic superpotentials are found to 
have a linear harmonic 
oscillator behavior with superposed 
oscillations consisting of several 
systematically varying frequencies. At the origin, cyclic 
superpotentials vanish when the period 
$p$ is odd, but diverge for $p$ even. The eigenvalue spectrum consists of 
$p$ infinite sets of equally spaced energy levels, shifted with respect 
to each other by arbitrary energies $\omega_0,\omega_1,\ldots,\omega_{p-1}$. 
As a special application, the 
energy spacings $\omega_k$ can be identified with the periodic points 
generated by the logistic map  $z_{k+1}=r z_k (1 - z_k)$. 
Increasing the value of $r$ and following the bifurcation 
route to chaos corresponds to studying cyclic shape invariant potentials 
as the period $p$ takes values $1,2,4,8,\ldots$
\end{abstract}

\newpage 

The number of analytically solvable eigenvalue problems in quantum 
mechanics is rather limited. In this regard, the 
concept of shape invariance \cite{Gendenshtein83} in supersymmetric 
quantum mechanics \cite{Witten81,Cooper95} 
has proved to be very useful because it leads immediately to exactly 
solvable potentials. 
In this paper, we define and study cyclic shape invariant 
potentials. The supersymmetric partner Hamiltonians correspond to 
a series of shape invariant potentials which repeats 
after a cycle of $p$ iterations. Such 
potentials have 
an infinite number of periodically spaced eigenvalues. More precisely, the 
level spacings are given by 
$\omega_0,\omega_1,\ldots,\omega_{p-1},
\omega_0,\omega_1,\ldots,\omega_{p-1},\omega_0,\omega_1,\ldots$, (see 
Fig. 1). The ground state is at zero energy; the next $(p-1)$ eigenvalues 
are $E_l = \sum_{k=0}^{l} \omega_k~,~l=0,1,\ldots,(p-2)~ $, and all other 
eigenvalues are obtained by adding arbitrary multiples 
of the quantity 
$\Omega_p \equiv \omega_0+\omega_1+\cdots+\omega_{p-1}$. The general 
formula for the excited energy levels is 
\beq \label{ev}
n \Omega_p + \sum_{k=0}^l \omega_k ~~;
~\{n=0,1,2,\ldots,\infty~~;~~l=0,1,\ldots,(p-1)\}~~.
\eeq
Effectively, the spectrum is obtained by starting 
with $p$ infinite sets of energy levels with equal spacing $\Omega_p$ 
and shifting 
these sets by successive arbitrary amounts 
$\omega_0,\omega_1,\ldots,\omega_{p-1}$. The superpotentials 
$W^{(p)}(x)$ corresponding to 
cyclic shape invariant potentials turn out to have rather interesting 
behavior at large $x$. One finds a linear dependence $\Omega_px/{2p}$ 
corresponding to a harmonic oscillator of angular frequency $\Omega_p/p$ with 
superposed sinusoidal oscillations which become more 
closely spaced with increasing 
$x$. We find that these oscillations at large $x$ are given 
by the functional form 
$b_{pj}\sin(q_{pj} x^2)$. For any choice of the period $p$, there are 
several oscillation 
``frequencies" $q_{pj}$, with $j=1,2,\ldots,[p]/2$, where  $[p]$ is the largest 
even integer less than $p$. These ``frequencies" are 
given by $q_{pj}=\Omega_p \tan(\pi j/p)/{2p}$. The amplitudes of 
the oscillations $b_{pj}$ 
depend on the energy spacings $\omega_0, \omega_1,\ldots$. 
The superpotentials also 
have unusual behavior at small $x$. For even values of $p$, one finds that 
$W^{(p)}(x)$ diverges like $1/x$,
whereas for odd values of $p$, $W^{(p)}(0)=0$. Our  plan is to 
first review some 
basic ideas of supersymmetric quantum mechanics and the concept of 
shape invariance for a change of parameters $a_1=f(a_0)$. Next, we formulate 
the general set of coupled nonlinear differential equations 
governing cyclic shape invariant potentials of arbitrary period $p$.
We describe some 
real transformations $f$ which have the property that $f^p(a_0)=a_0$, that 
is the parameters $a_k~,~k=0,1,\ldots,(p-1)$ repeat with 
period $p$. We study and discuss the 
solutions of the cyclic shape invariance conditions. The
special cases of periods $p=3,4$ are treated in some detail. 
Many properties generalize to arbitrary values of $p$. Numerical solutions 
for $W^{(p)}(x)$ and the corresponding potentials $V_-^{(p)}(x)$ 
illustrate the general results. As an application, we examine cyclic 
shape invariant potentials corresponding to energy level spacings 
$\omega_k$ which are the periodic points 
generated by the logistic map  $z_{k+1}=r z_k (1 - z_k)$ \cite{Peitgen92}. 
As the parameter $r$ increases from $1$ to $3.569946$, 
the period $p$ also increases and the energy level spacings become increasingly 
more chaotic. This is reflected in the fact that the potentials $V_-(x)$ 
show increasingly more oscillation frequencies as one follows the 
period doubling route toward chaos. \sss

\noindent {\bf General Formalism for Cyclic Shape Invariant Potentials:}\sms

Recall that the superpotential $W(x,a_0)$ generates the supersymmetric 
partner potentials 
\beq \label{vpm}
V_{\pm}(x,a_0)=W^2(x,a_0) \pm W'(x,a_0)~~,
\eeq
where $a_0$ is a set of parameters. These partner potentials are 
shape invariant if they both have the same $x$-dependence upto a change 
of parameters $a_1=f(a_0)$ and an additive constant $R(a_0)$. The 
shape invariance condition is \cite{Gendenshtein83}
$~V_+(x,a_0) = V_-(x,a_1) + R(a_0)~$,
or equivalently
\begin{equation} \label{sipw}
W^2(x,a_0)+W'(x,a_0) = W^2(x,a_1)-W'(x,a_1)+R(a_0)~~.
\end{equation}
The property of shape invariance permits an immediate analytic 
determination of energy eigenvalues \cite{Gendenshtein83}, eigenfunctions 
\cite{Dutt86} and scattering matrices \cite{Khare88}. For unbroken 
supersymmetry, the eigenstates of the potential $V_-(x)$ are:
$$
E_0^{(-)} =0~,~E_n^{(-)}=\sum_{k=0}^{n-1} R(a_k)~,
$$
\begin{equation}
\psi_0^{(-)} \propto e^{- \int^x_{x_0} W(y,a_0) dy}~,~
\psi_n^{(-)}(x,a_0)=\left[-\frac{d}{dx}+W(x,a_0)\right]
\psi_{n-1}^{(-)}(x,a_1)~~(n=1,2,3,\ldots)~.
\label{psi}
\end{equation}
The lowest lying eigenstate is at zero energy, characteristic 
of unbroken supersymmetry.

Solutions of the shape 
invariance condition (\ref{sipw})
corresponding to a translational change of parameters
$a_1=f(a_0)=a_0+\beta~,~a_n=f^n(a_0)=a_0+n \beta~$ 
are well-known \cite{Cooper95}. Also, 
the scaling change of parameters
$a_1=f(a_0)=q a_0~,~a_n=f^n(a_0)=q^n a_0$
has been investigated and shown to produce new classes of shape invariant 
potentials \cite{Khare93,Barclay93}, which included self-similar 
potentials \cite{Shabat92} 
as a special case. 

Very recently, another type of change of parameters defined by 
$f^2(a_0)=a_0$ was studied in detail \cite{Gangopadhyaya96}. Here, the 
parameters repeat after a cycle of two iterations and the spectrum consists 
of two shifted sets of equally spaced eigenvalues. In this 
paper, we consider a much broader change of parameters defined 
by 
\beq \label{fp}
f^p(a_0)=a_0,
\eeq 
where the 
parameters will repeat after a cycle of $p$ iterations. For 
this case, one has
$a_0=a_p~~, ~~f(a_0)=a_1=a_{p+1}$, etc.

For a cyclic set of parameters, we now formulate the corresponding 
shape invariance conditions (\ref{sipw}) for any period $p$. Defining 
$R(a_k) \equiv \omega_k$ for 
$k=0,1,\ldots,(p-1)$, and using $a_p=a_0$, one obtains a set of 
$p$ coupled nonlinear 
differential equations
\beq \label{coupled}
W^2(x,a_k)+W'(x,a_k) = W^2(x,a_{k+1})-W'(x,a_{k+1})+\omega_k~~,~~
k=0,1,\ldots,(p-1)~.
\eeq
There are $p$ unknown superpotentials $W(x,a_k)$ which are related 
to each other by cyclic permutations of the indices $0,1,2,\ldots,(p-1)$. 
We are seeking superpotentials with odd parity, which will yield 
symmetric potentials $V_-(x)$. It is convenient to 
denote $W(x,a_k) \equiv W_k$. The set of shape invariance conditions now has
a deceptively succinct form:
\beq \label{main}
W^2_k+W'_k=W^2_{k+1} - W'_{k+1}+\omega_k~~,~k=0,1,\ldots,(p-1)~.
\eeq

There are two important problems one needs to solve: 
(i) to find real transformation functions 
$f(y)$ which satisfy eq. (\ref{fp}) and hence produce a cyclic 
set of parameters ; (ii) to determine the superpotentials 
$W_k^{(p)}(x)$ by solving 
the cyclic shape invariance conditions (\ref{main}) 
corresponding to an arbitrary cyclic 
choice of parameters. These two problems will be treated in turn.\sss

\noindent {\bf Transformations Giving Cyclic Sets of Parameters:}\sms

We now show that 
a simple algorithm for generating non-trivial real functions
$ f : {\rm {\bf R}} \to {\rm {\bf R}} $ satisfying $f^p(y)=y~,~
\forall y \in {\rm {\bf R}}$ is the projective(Moebius) transformation 
\beq \label{proj}
f(y)=\frac{ay+b}{cy+d}~~,
\eeq
with specific constraints on the parameters $a,b,c,d$. Let $M$ be the 
real $2 \times 2$ matrix 
\be
M=\fourmat{a}{b}{c}{d}~.
\lab{matr}
\ee
For $f(y)$ given by eq. (\ref{proj}), observe that $f^p(y)$ is also a 
projective transformation whose coefficients are the same as the elements 
of the matrix $M^p$.
Therefore, solving eq. (\ref{fp}) can be rephrased in terms of finding
a real $2 \times 2$ matrix $M$, such that for any integer $p \ge 1$, the 
matrix $M^p$ is proportional to the unit matrix ${\bf 1}$. Without loss 
of generality, we choose the condition
$M^p = {\bf 1}$, and our problem is to find 
real $2$-dimensional representations of the cyclic group of
order $p$. This group being Abelian, its $2$-dimensional representations
are reducible, i.e. there exists a non-singular matrix $Q$ for 
all matrices $M$ such that 
\be
\cM \equiv Q^{-1} M Q = \fourmat{\l_1}{0}{0}{\l_2}~.
\lab{diag}
\ee
Here $\l_k$ ($k=1,2)$ stands for any two of the $p$-th roots of unity:
\be
\l_k = \exp \( \frac{2 \pi j_k}{p}\, i\)~,
          \quad \quad j_k = 0,\ldots, p-1~.
\lab{roots}
\ee
Since we are interested in non-trivial matrices, we will impose two additional
constraints: $\l_1 \neq \l_2$ and $M^n \neq {\bf 1}$
for any $n<p$.
The range of possible solutions is constrained by 
the invariance under similarity 
transformations \rf{diag} of both the trace and 
the determinant:
\beq
a + d     = \exp \( \frac{2 \pi j_1}{p}\, i\) 
            + \exp \( \frac{2 \pi j_2}{p}\, i\) ~;~~
a d - b c = \exp \( \frac{2 \pi (j_1+j_2)}{p}\, i\)~.
\eeq
Also, the reality condition for the elements of $M$ gives
\beq \label{real}
\sin \( \frac{2 \pi j_1}{p} \) +\sin \( \frac{2 \pi j_2}{p}\)=0~;~
\sin \( \frac{2\pi (j_1 + j_2)}{p} \) = 0~.
\eeq
Taking into account all the above considerations we are left only with
the roots of unity  $\l_1$ and $\l_2$ satisfying $j_1+j_2 = p$.
Therefore, the matrix ${\cal M}$ will be a matrix of the form \rf{diag}
with $\lambda_1= \exp \( \frac{2 \pi j_1}{p}\, i\) = \lambda_2^{-1}$. Note 
that the choice $j=0$ gives ${\cal M}={\bf 1}$, which gives a trivial $M$. 
Similarly, if $p$ is a non-prime number, say $p=jn$, then 
${\cal M}^n={\bf 1}$ for $n < p$, which gives 
$M^n={\bf 1}$, and is not a non-trivial 
result for $M$. However, for all prime periods $p$, the matrix $M$ has the 
form \rf{matr} whose elements must be real and 
satisfy the system of equations:
\beq \label{sys}
a d - b c = 1~;~
a + d = 2 \cos \( \frac{2 \pi j}{p} \)~
       \quad, \quad \quad \(j=0,1, \ldots, p-1\)~.
\eeq

Finally, the desired transformation 
function $f(y)$ is given by eq. (\ref{proj}) with real 
parameters $a,b,c$ and $d$ satisfying the constraints (\ref{sys}).
For example, a projective transformation 
corresponding to the choices $p=3, j=1$ is
$$
f(y)=\frac{ay+b}{-y(a^2+a+1)/b-(a+1)}~~,
$$
and it is easy to check that $f^3(y)=y$.
\sss

\noindent {\bf Solutions of the Cyclic Shape Invariance Conditions:}\sms

Having established that there exist real transformations $f$ which 
produce cyclic sets of parameters, we now turn our attention to 
studying the solutions of the cyclic set of equations (\ref{main}). A very 
general result comes from 
adding all the equations. It gives $\sum_{k=0}^{p-1}W'_k=\Omega_p/2$, 
which on integration yields 
\beq \label{sum}
\sum_{k=0}^{p-1} W_k(x)=\frac{1}{2} \Omega_p x~~.
\eeq
This condition is valid for all $x$. There is no constant of integration 
in eq. (\ref{sum}) since we want the superpotentials $W_k$ to be odd 
functions of $x$. In particular, at large $x$, constraints (\ref{main}) and 
(\ref{sum}) give all $W_k$ the linear asymptotic behaviour $\Omega_p x/2p$, 
corresponding to a simple harmonic oscillator potential of angular 
frequency $\Omega_p/p$. Also, at small $x$, in order to satisfy eqs. 
(\ref{main}), all superpotentials can at most have a $1/x$ singularity.

The general analytic solution of equations (\ref{main}) is not easy to obtain. 
However, a numerical solution can be obtained via the Runge-Kutta method 
and we will also obtain several analytic results at small 
and large values of $x$. For ease of presentation, we systematically consider 
small values of the cycle period $p$ before coming to general results. \sss

\noindent {\bf $p=1$:}   $~~$ Here $a_1=f(a_0)=a_0$ and all energy levels are 
equally spaced. As expected, the shape invariance condition (\ref{main}) 
gives $W_0^{(1)}=\omega_0 x/2$ which is a simple 
harmonic oscillator superpotential. \sss

\noindent {\bf $p=2$:}  $~~$ This is the case of two shifted sets of equally 
spaced eigenvalues considered in detail in ref. \cite{Gangopadhyaya96}. 
The shape invariance conditions
(\ref{main}) are:
\begin{equation} \label{eq9}
W^2_0+W'_0 = W^2_1-W'_1+\omega_0~~~,~~
W^2_1+W'_1 = W^2_0-W'_0+\omega_1~~.
\end{equation}
The solution is obtained by straightforward manipulations.
The superpotential $W_0^{(2)} \equiv W^{(2)}(x,a_0)$ 
and the corresponding potential 
$V_-^{(2)}(x)$  obtained using eq. (\ref{vpm}) are given by 
\begin{equation}
W_0^{(2)}(x)=\frac{\Omega_2 x}{4}+\frac{(\omega_0-\omega_1)}
{2 \Omega_2 x}~,
~V_-^{(2)}(x)=
\frac{(\omega_0-\omega_1)(3\omega_0+\omega_1)}
{4{\Omega_2}^2 x^2} 
- \frac{\omega_1}{2}
+\frac{{\Omega_2}^2 x^2}{16}~.
\label{potential}
\end{equation}
Note that for period $p=2$, the superpotential $W_0^{(2)}$ 
diverges like $\alpha_0/x$ near the origin, which produces a 
singular inverse square potential in the transition 
region \cite{Landau77,Frank71,Gangopadhyaya94}. A detailed 
discussion of the potential as well as the eigenfunctions $\psi_n^{(-)}$ 
obtained from 
eq. (\ref{psi}) is given in ref. \cite{Gangopadhyaya96}. 
\sss

\noindent {\bf $p=3$:} $~~$ There are now three shape invariance 
conditions (\ref{main}):
\beq \label{eq3}
W^2_0+W'_0 = W^2_1-W'_1+\omega_0~,~
W^2_1+W'_1 = W^2_2-W'_2+\omega_1~,~ 
W^2_2+W'_2 = W^2_0-W'_0+\omega_2~.
\eeq
These equations are not easy to solve analytically. In order to 
obtain a numerical solution, it is best to first solve eqs. (\ref{eq3}) 
for the derivatives $W'_0,W'_1,W'_2$. One gets
\br \label{eq3rk}
W'_0 &=& W^2_1-W^2_2+\frac{1}{2}(\omega_0-\omega_1+\omega_2)~~, \nonu \\
W'_1 &=& W^2_2-W^2_0+\frac{1}{2}(\omega_1-\omega_2+\omega_0)~~, \nonu \\
W'_2 &=& W^2_0-W^2_1+\frac{1}{2}(\omega_2-\omega_0+\omega_1)~~.
\er
In this form the Runge-Kutta method is immediately applicable, and 
the superpotentials can be determined for any choice of the energy 
spacings $\omega_0,\omega_1,\omega_2$. An illustration of the 
results is shown in Fig. 2. The 
overall average behavior of 
$W_0^{(3)}$ at large $x$ is indeed seen to be 
$\Omega_3 x/6$ as expected. On this linear dependence are clear oscillations 
which get more closely spaced as $x$ increases. They correspond to a 
behavior $b \sin (qx^2)$ where the amplitude $b$ and the 
frequency $q$ both depend on the 
choice of energy spacings 
$\omega_0,\omega_1,\omega_2$. In particular, when one chooses 
$\omega_0=\omega_1=\omega_2$,
the numerical solution shows no oscillations. This is reasonable since 
for this case one expects a simple harmonic superpotential $\Omega_3 x/6$.

One can get a number of analytic results for 
eqs. (\ref{eq3rk}) at small $x$ as well as large $x$. At 
small $x$, if one feeds the series expansions 
\beq \label{wseries}
W_k=\alpha_k/x+\beta_k x +\gamma_k x^3 + \delta_k x^5 + \cdots
\eeq
for $k=0,1,2$ into eq. (\ref{eq3rk}) and equates coefficients of $x$, one gets 
\beq
W_0^{(3)}=\frac{x}{2}(\omega_0-\omega_1+\omega_2) 
+\frac{x^3}{3} \omega_1 (\omega_0-\omega_2)+\frac{x^5}{15}\omega_1
(\omega_0 \omega_1+\omega_1 \omega_2 - \omega_0^2 - \omega_2^2)+\cdots
\eeq
with $W_1^{(3)}, W_2^{(3)}$ obtained by cyclic permutations of the indices 
$0,1,2$. $W_0^{(3)}$ 
vanishes at $x=0$. Note that 
$\sum_{k=0}^2 \gamma_k = \sum_{k=0}^2 \delta_k= 0$ in 
agreement with eq. (\ref{sum}).
At large $x$, one finds oscillations with one 
frequency $q_{31}=\Omega_3 \sqrt{3}/6$. 
This is a special case of the result for arbitrary $p$ which will be
derived below. 
\sss 

\noindent {\bf $p=4$:} $~~$ It is worth discussing period $4$ briefly, 
since several results are different 
from period 3. There are four cyclic shape invariance conditions:
\beqr  \label{eq4}
W^2_0+W'_0 &=& W^2_1-W'_1+\omega_0~,~
W^2_1+W'_1 = W^2_2-W'_2+\omega_1~,~\nonu \\
W^2_2+W'_2 &=& W^2_3-W'_3+\omega_2~,~ 
W^2_3+W'_3 = W^2_0-W'_0+\omega_3~.
\eeqr
Unlike the period $3$ case, one cannot directly solve for 
$W'_0,W'_1,W'_2,W'_3$, since their coefficients in eqs. (\ref{eq4}) 
have a zero determinant. Indeed, this situation occurs for all even 
periods $p$. For the case of period 4, one can proceed by noting that 
eqs. (\ref{eq4}) give the constraint 
$W^2_0-W^2_1+W^2_2-W^2_3=(\om_0-\om_1+\om_2-\om_3)/2$, 
which on differentiation yields
$$W_0W'_0-W_1W'_1+W_2W'_2-W_3W'_3=0.$$ This equation, along with any three of 
eqs. (\ref{eq4}), permits one to solve for 
$W'_0,W'_1,W'_2,W'_3$, and proceed with a numerical determination of the 
superpotentials by the Runge-Kutta method. 

Analytically, at small $x$, substitution of expansions 
of the form (\ref{wseries}) into eqs. (\ref{eq4}) and collecting powers 
of $x$ gives 
\beq
W_0^{(4)}=\frac{(\om_0-\om_1+\om_2-\om_3)}{2 \Omega_4 x}
+\frac{x \Omega_4 (\om_0\om_3+\om_0\om_1+\om_2\om_3-\om_2\om_1)}
 {4(\om_0+\om_2)(\om_1+\om_3)}+\cdots
\eeq
with the other superpotentials $W_1^{(4)},W_2^{(4)},W_3^{(4)}$
obtained by cyclic permutations of the indices
$0,1,2,3$. Note the singular $1/x$ dependence, similar to period $2$. This 
is a feature of all even periods.
At large $x$, period $4$ superpotentials show oscillations with a single 
frequency $q_{41}=\Omega_4/8$. 
We will derive the general result for arbitrary $p$ below.
\sss

\noindent {\bf Results for Arbitrary $p$:} $~~$ The experience 
gained from small values of the period $p$ can be used 
to derive a number of general analytic statements valid for 
any arbitrary period $p$.
At small $x$, the superpotential $W_0^{(p)}(x)$ has the form
\br
W_0^{(p)}&=& \frac{x}{2}\( \om_0-\om_1+\cdots+\om_{p-1}\) \nonu \\
 &-& \frac{x^3}{3} \Blb \,\om_1\(\om_2-\om_3+\cdots+\om_{p-1}-\om_0 \) +
  \om_3\(\om_4-\om_5+\cdots+\om_{p-1}-\om_0+\om_1-\om_2 \) \nonu \\
 &+& \! \! \! \! \cdots+\om_{p-2}\( \om_{p-1}-\om_0+\om_1-\cdots-\om_{p-3}\) 
\,\Brb
  +\;\, O(x^5)~; \qquad (p~~ {\rm odd}) \nonu \\
W_0^{(p)}&=&\frac{1}{2 \Omega_p x} 
\( \om_0-\om_1+\om_2-\cdots-\om_{p-1}\) +\frac{x\Omega_p}{4}\, 
\Blb \,1 - 2\times \nonu \\
& \times & \! \! \! 
\frac{\om_1\(\om_2+\om_4+\cdots+\om_{p-2}\)+\om_3 \(\om_4+\om_6+\cdots
+\om_{p-2}\) + \cdots + \om_{p-3}\om_{p-2}}
{\(\om_0+\om_2+\cdots+\om_{p-2}\)\(\om_1+\om_3+\cdots+\om_{p-1}\)}\, 
\Brb \nonu \\
&+&O(x^3)~; \qquad (p~~ {\rm even})
\er
The behaviour of the other 
superpotentials $W_k^{(p)}(x)$
can be seen by cyclically permuting the indices $0,1,2,\ldots,(p-1)$.
Note the $1/x$ singularity at the origin for even periods. As for the 
case of periods $p=2,4,$ it can be readily verified that the coefficient of the 
$1/x^2$ singularity in the potential $V_k^{(p)}$ lies in the transition 
region $-1/4$ to $3/4$. In this region, both solutions of the Schr{\" o}dinger 
equation are square integrable at 
the origin \cite{Landau77,Frank71,Gangopadhyaya94}. The infinite discontinuity 
in $W^{(p)}(x)$ at the origin for even $p$ produces a 
$\delta(x)$ term in the potential 
$V_-^{(p)}(x)$ as described in ref. (\cite{Gangopadhyaya96}).

At large $x$, one removes the asymptotic linear behaviour by defining
\beq \label{qp}
Q_k(x) \equiv W_k(x)-\Omega_p x/2p~,~~k=0,1,\ldots,p-1~~.
\eeq
Since numerical solutions suggest sinusoidal oscillations in $x^2$, we 
make a change of variables to $u \equiv x^2$. In terms of $u$, the equations 
satisfied by $Q_k$ are obtained by substituting eq. (\ref{qp}) 
into (\ref{main}). Retaining only the leading powers of $u$ yields
\beq
Q'_0 + Q'_1 = \frac{\Omega_p}{2p}\(-Q_0+Q_1\)
\eeq
and cyclic permutations. This set of equations can be written in a 
more compact form by putting all the superpotentials into a column 
vector $Q$ and making use of the matrix $C_p$ whose only non-zero elements 
are $(C_p)_{12} =(C_p)_{23} = \cdots =(C_p)_{p-1,p} =(C_p)_{p1} = 1 $.
The equations now read
\beq
(C_p+1)Q'-\frac{\Omega_p}{2p} (C_p-1)Q = 0~~.
\eeq
The most general solution is a linear combination of eigenstates of the 
form $Q \propto \exp{(\lambda u)}$. The 
eigenvalues $\lambda$ are determined from
\beq \label{ll}
\det [\lambda(C_p+1)-(\Omega_p/2p)(C_p-1)] = 0~~.
\eeq
Defining $\tilde \lambda = 2p \lambda/ \Omega_p$,
and expanding the determinant, eq. (\ref{ll}) can be re-written as
$$ (\tilde\lambda+1)^p-(-1)^p (\tilde \lambda -1)^p=0~~,$$
which is a polynomial equation of degree $p$ for $p$ odd and of degree 
$p-1$ for $p$ even. This situation occurs because eq. (\ref{ll}) is more 
general than the standard eigenvalue equation. The solution is
$$ \frac{1-\tilde \lambda}{1+\tilde \lambda}=
\exp \( \frac{2 \pi j}{p}\, i\)~,
~~j= 0, \pm 1,\pm 2,\cdots,\pm [p]/2~~.$$
$[p]$ is the largest even integer less than $p$. This gives rise 
to distinct oscillation frequencies 
\beq \label{freq}
q_{pj}=(\Omega_p/2p) \tan (\pi j/p)~~,~~ 
j=1,2,\ldots,[p]/2~~.  
\eeq

At large $x$, the 
superpotential $W_0^{(p)}(x)$ has the form 
\beq
W_k^{(p)}(x) = \frac{\Omega_p x}{2 p}
+ \sum _j b_{pj} \sin(q_{pj} x^2+\phi_j^{(k)})+O(1/x)
\eeq
where the phases are 
\be
\phi_j^{(k)}=\phi_j+2\pi k/p~~.
\ee
Note that the requirement of having odd superpotentials can be satisfied by 
inserting an asymptotically constant odd function such as 
$\tanh (\alpha x)$ or $\tan^{-1} (\alpha x)$ 
with the coefficient $b_{pj}$. 
This does not affect the above discussion of oscillation frequencies.
Also, if one uses eq. (\ref{psi}) to get the asymptotic behaviour of 
wave functions, the linear term in $W^{(p)}$ gives the familiar Gaussian 
dependence of a harmonic oscillator and the sinusoidal correction terms 
in $W^{(p)}$ give Fresnel integrals producing oscillations on the Gaussian.

As an example, for period $5$ , taking $\Omega_5=6.5$, we predict two 
oscillation frequencies $q_{51}=0.472$ 
and $q_{52}=2.001$ from eq. (\ref{freq}), and we have checked that these 
values are in agreement with the 
numerical solutions shown in Fig. 3.
\sss

\noindent {\bf Approach to Quantum Chaos Using the Logistic Map:}\sms

We have now established the method of generating cyclic shape invariant 
potentials. This method can be applied using 
any transformation $f(y)$ which generates 
parameters which repeat with period $p$. 
For the special projective transformations 
described previously, the initial parameter $a_0$ can be chosen arbitrarily. 
There are also other possible transformations in which cyclic parameters 
are obtained if one starts from specific choices for $a_0$. One such 
interesting possibility, which we discuss in detail here, is to use a set of 
cyclic parameters given by the periodic points of the logistic map
\beq \label{logistic}
z_{k+1}=r z_k (1 - z_k)~~ .
\eeq
This famous transformation has been extensively 
studied \cite{Peitgen92}. For $1<r<3$, one has a single non-zero fixed point 
at $z=(r-1)/r$; for $3<r<3.449499$, there is a stable 2-cycle 
with periodic points at $z=[(r+1) \pm \sqrt {(r-3)(r+1)}]/2r$; 
for $3.449499<r<3.544090$, there is a 
stable 4-cycle, etc.  As $r$ is increased, the period keeps 
doubling, giving finally the period $2^\infty$ at $3.569946$. For 
$r>3.569946$, one obtains all odd periods and their harmonics, and at $r>4$,
there is complete chaos.

In our application, we will identify the energy spacings $\omega_k$ with 
the periodic points of the logistic map. For example, if we choose 
$r=3.20$, the periodic points are $0.799$ and $0.513$, and we 
determine the potentials which have energy spacings $\omega_0=0.799$ and 
$\omega_1=0.513$. The corresponding superpotential is shown in Fig 4(a). 
Other examples with larger values of $r$ and correspondingly larger 
periods $p=4,8$ are also shown in Figs. 4(b) and 4(c).  
Clearly, as one advances along the period doubling route to chaos, more and 
more oscillation frequencies come into play - which is certainly a 
plausible result. Specifically, for period $2^n$, there are 
$2^{n-1}-1$ frequencies given by eq. (\ref{freq}). 

Although we have so far chosen values of $r$ which yield stable cycles 
of period $2^n$, it is known that there are 
small windows of $r$-values where stable cycles of periods 3,5,\ldots also 
exist. The above methods immediately permit us to obtain superpotentials 
for these cases also. These issues and deeper connections with chaotically 
spaced energy levels will be studied further. Parenthetically, we note that the 
locations of the nontrivial zeros of the 
Riemann zeta function appear to exhibit chaotic behaviour. If these positions
are identified with energy level spacings, the corresponding 
potentials have many 
oscillations \cite{Wu93}.
\sss

We would like to thank A. Gangopadhyaya and T. Imbo for valuable discussions 
and comments. U.S. acknowledges the hospitality of the Institute of Physics, 
Bhubaneswar where part of this work was done. This research was supported 
in part by the U.S. Department of Energy.

\newpage

{\bf \Large Figure Captions}\sss

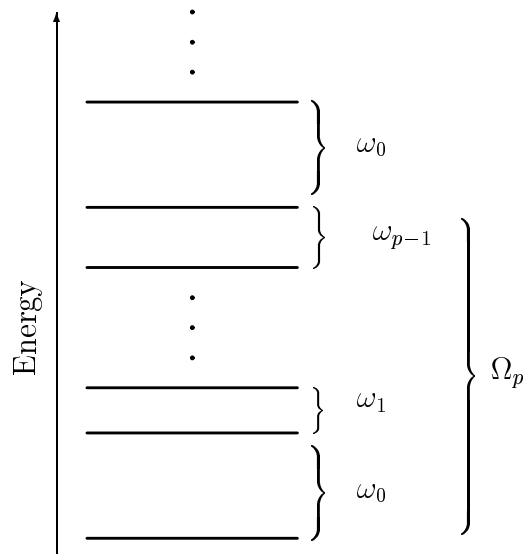
{\bf Figure 1:}  This diagram illustrates the energy levels of a cyclic 
shape invariant potential of period $p$. The lowest level is at zero 
energy, and the other excited energy levels are given by (\ref{ev}).\sss

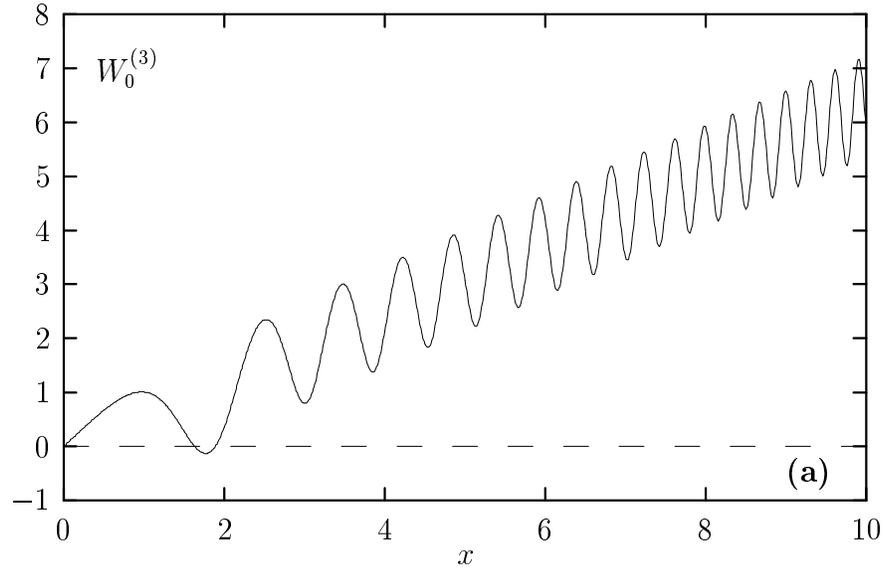
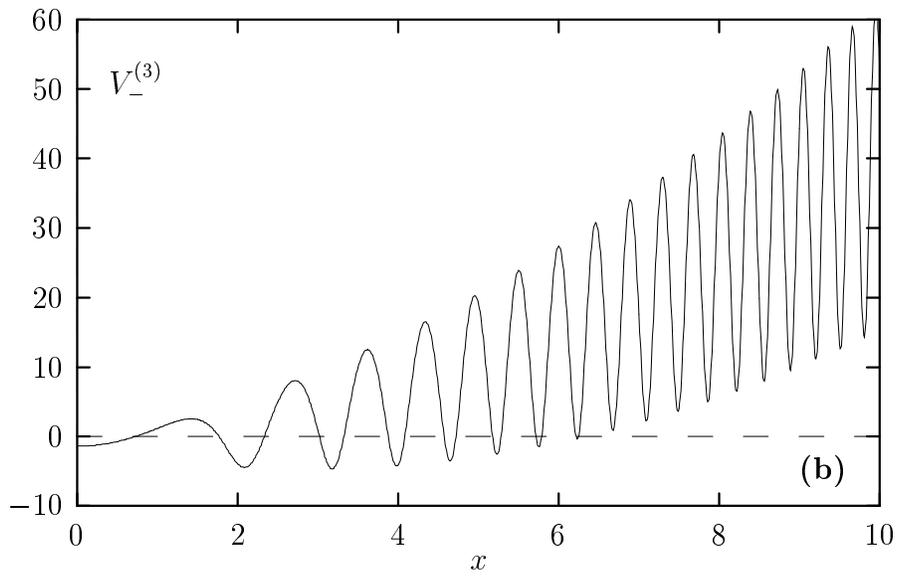
{\bf Figure 2:}  The superpotential $W_0^{(3)}(x)$ and the corresponding 
symmetric potential $V_{-}^{(3)}(x)$ for period $p=3$ and 
a choice of energy level
spacings $\omega_0=1.0, ~\omega_1=0.5, ~\omega_2=2.3 $. The superpotential
was obtained by a numerical solution of eqs. (\ref{eq3rk}) by 
the Runge-Kutta method. The oscillation frequency given by  eq. (\ref{freq}) 
is $q_{31}=1.097$, and is in agreement with the figure. This is established 
by counting the number of maxima in any chosen interval of $x$ values.
The energy levels for this potential are given by expression (\ref{ev}) 
and are located at 
$n\omega_3, n\omega_3 + 1.0, n\omega_3+1.5$ with $\omega_3=3.8$ 
and $n=0,1,2,\ldots$  \sss

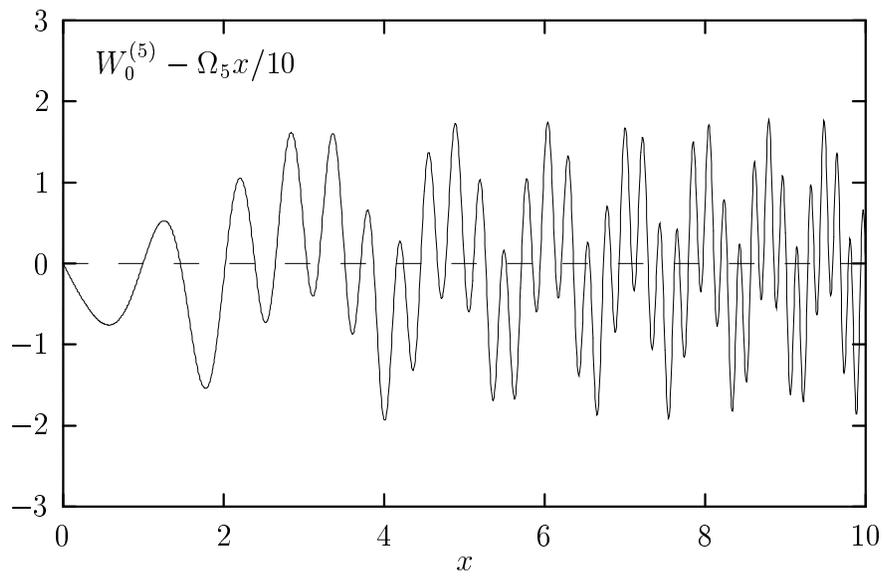
{\bf Figure 3:}  A plot of the superpotential $W_0^{(5)}-\frac{\Omega_5 x}{10}$
for period $p=5$ and energy level spacings 
$\omega_0=0.8,~\omega_1=1.5,~\omega_2=1.0,~\omega_3=3.0,~\omega_4=0.2$ .
The asymptotic behavior $\frac{\Omega_5 x}{10}$ has been subtracted in
order to clearly reveal the two oscillation frequencies $q_{51}=0.4722$ and
$q_{52}=2.001$, predicted by eq. (\ref{freq}). The maxima are spaced 
in agreement with the smaller frequency $q_{51}$ whereas their 
``envelope" corresponds to the larger frequency $q_{52}$. \sss

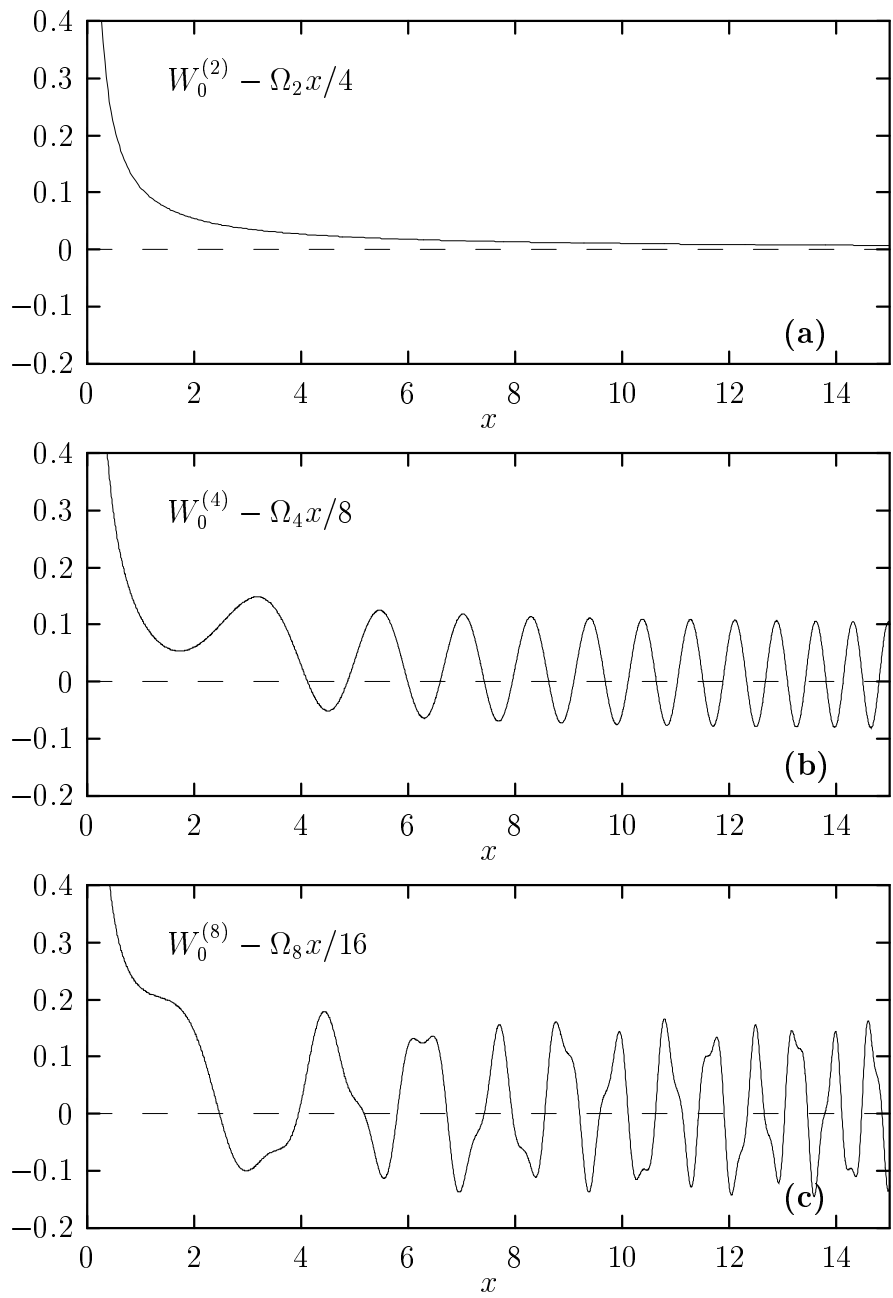
{\bf Figure 4:}  This figure shows how the superpotential 
$W_0^{(p)}(x)-\frac{\Omega_p x}{2p}$ gets progressively more complicated 
and aquires more oscillation frequencies as the period $p$ is successively 
doubled along the bifurcation route to chaos via the logistic 
map (\ref{logistic}). 
The parameter values chosen for the figures are:

        (a) $r=3.200$ which gives period $p=2$ and a stable 2-cycle with
energy spacings  $\omega_0=0.799,~\omega_1=0.513$~.

        (b) $r=3.500$ which gives period $p=4$ and a stable 4-cycle with
$\omega_0=0.827, ~\omega_1=0.501, ~\omega_2=0.875, ~\omega_3=0.383$. The 
oscillation frequency from eq. (\ref{freq}) is $q_{41}=0.323$.

        (c) $r=3.561$ which gives period $p=8$ and a stable 8-cycle with
$\omega_0=0.890, ~\omega_1=0.348, ~\omega_2=0.808, ~\omega_3=0.552,
~\omega_4=0.881, ~\omega_5=0.374, ~\omega_6=0.834, ~\omega_7=0.494 $. The 
three oscillation frequencies given by eq. (\ref{freq}) are 
$q_{81}=0.1341, ~q_{82}=0.3238, ~q_{83}=0.7818$.

\end{document}